\documentclass[twocolumn,english,aps,prd,graphics,floatfix,a4paper,superscriptaddress]{revtex4-1} 
\usepackage{amsmath, amssymb}
\usepackage{makecell}
\usepackage{multirow}
\usepackage{CJK}
\usepackage{graphicx}
\usepackage{mathrsfs}
\usepackage{bm}
\usepackage{amsmath}
\usepackage{dcolumn}
\usepackage{epstopdf}
\usepackage{dsfont}
\usepackage{amssymb}
\usepackage{tabularx}
\usepackage{array}
\usepackage{float}
\usepackage{color}
\usepackage{epstopdf}
\usepackage{mathrsfs}

\usepackage[colorlinks, linkcolor=blue,anchorcolor=blue,citecolor=blue,urlcolor=blue]{hyperref}
\usepackage{extarrows}

\begin{document}
\title{Magnetic eight-fold nodal-point and nodal-network fermions in MnB$_{2}$}

\author{Yongheng Ge}
\affiliation{Key Laboratory of Low-Dimensional Quantum Structures and Quantum Control, Ministry of Education, Department of Physics and Synergetic Innovation Center for Quantum Effects and Applications, Hunan Normal University, Changsha 410081, China}	

\author{Ziming Zhu}\email{zimingzhu@hunnu.edu.cn}
\affiliation{Key Laboratory of Low-Dimensional Quantum Structures and Quantum Control, Ministry of Education, Department of Physics and Synergetic Innovation Center for Quantum Effects and Applications, Hunan Normal University, Changsha 410081, China}

\author{Zeying Zhang}\email{zzy@mail.buct.edu.cn}
\affiliation{College of Mathematics and Physics, Beijing University of Chemical Technology, Beijing 100029, China}
\affiliation{Research Laboratory for Quantum Materials, Singapore University of Technology and Design, Singapore 487372, Singapore}	

\author{Yi Shen}
\affiliation{Key Laboratory of Low-Dimensional Quantum Structures and Quantum Control, Ministry of Education, Department of Physics and Synergetic Innovation Center for Quantum Effects and Applications, Hunan Normal University, Changsha 410081, China}	

\author{Weikang Wu}
\affiliation{Key Laboratory for Liquid-Solid Structural Evolution and Processing of Materials, Ministry of Education, Shandong University, Jinan 250061, China}
\affiliation{Research Laboratory for Quantum Materials, Singapore University of Technology and Design, Singapore 487372, Singapore}

\author{Cong Xiao}
\affiliation{Institute of Applied Physics and Materials Engineering, University of Macau, Macau SAR}

\author{Shengyuan A. Yang}
\affiliation{Institute of Applied Physics and Materials Engineering, University of Macau, Macau SAR}
\affiliation{Research Laboratory for Quantum Materials, Singapore University of Technology and Design, Singapore 487372, Singapore}

\begin{abstract}
Realizing topological semimetal states with novel emergent fermions in magnetic materials is a focus of current research. Based on first-principles calculations and symmetry analysis, we reveal interesting magnetic emergent fermions in an existing material MnB$_2$. In the temperature range from 157 K to 760 K, MnB$_2$ is a collinear antiferromagnet. We find the coexistence of eight-fold nodal points and nodal network close to the Fermi level, which are protected by the spin group in the absence of spin-orbit coupling.
Depending on the N\'{e}el vector orientation, consideration of spin-orbit coupling will either open small gaps at these nodal features, or transform them into magnetic linear and quadratic Dirac points and nodal rings. Below 157 K, MnB$_2$ acquires weak ferromagnetism due to spin tilting. We predict that this transition is accompanied by a drastic change in anomalous Hall response, from zero above 157 K to $\sim 200$ $\Omega\cdot \text{cm}^{-1}$ below 157 K.

\end{abstract}

\maketitle

Topological semimetals have been attracting great interest in condensed matter physics research~\cite{RevModPhys.88.035005,annurev-conmatphys-031016-025458,RevModPhys.90.015001,RevModPhys.93.025002}.
In these materials, the symmetry/topology protected band degeneracies in the
vicinity of Fermi level give rise to novel emergent fermion states, which may lead to fascinating physical effects.
Early examples include Weyl and Dirac semimetals, which host twofold and fourfold nodal points, respectively, and can simulate the physics of Weyl and Dirac fermions~\cite{RevModPhys.90.015001}. Subsequent works showed that there is a rich variety of emergent fermions
beyond the Weyl and Dirac paradigm, which may have different number of degeneracy and different dimension of the band degeneracy manifold~\cite{PhysRevLett.108.266802,yang2014classification,doi:10.1126/science.aaf5037,PhysRevLett.116.186402,PhysRevX.6.031003,YU2022375}.
Recent classification works showed that the maximal degree of degeneracy that can be protected by crystalline symmetry is eight~\cite{PhysRevLett.116.186402,doi:10.1126/science.aaf5037,YU2022375,PhysRevB.105.085117,PhysRevB.105.104426,PhysRevB.105.155156,rong2023realization}. Such eightfold nodal point (ENP) has been proposed in a few nonmagnetic crystals, such as  Bi$_2$CuO$_4$~\cite{doi:10.1126/science.aaf5037},Bi$_2$AuO$_5$~\cite{PhysRevLett.116.186402}, TaCo$_2$Te$_2$~\cite{rong2023realization} and etc.
In Refs.~\cite{rong2023realization}, ENP in TaCo$_2$Te$_2$ has been probed by angle-resolved photoemission spectroscopy (ARPES).
Besides nodal points which are zero-dimensional (0D), the degeneracy manifold may also form 1D nodal lines~\cite{PhysRevLett.113.046401,weng2015topological,fang2016topological,PhysRevB.107.205120}
 or even 2D nodal surfaces~\cite{liang2016node,zhong2016towards,PhysRevB.96.155105,wu2018nodal}
 . Especially, the nodal lines can be connected to form various patterns in the momentum space~\cite{bzduvsek2016nodal,wang2017hourglass,PhysRevLett.119.036401,sheng2017d}.


\begin{figure*}[htp!]
	\centerline{\includegraphics[width = 0.9\linewidth]{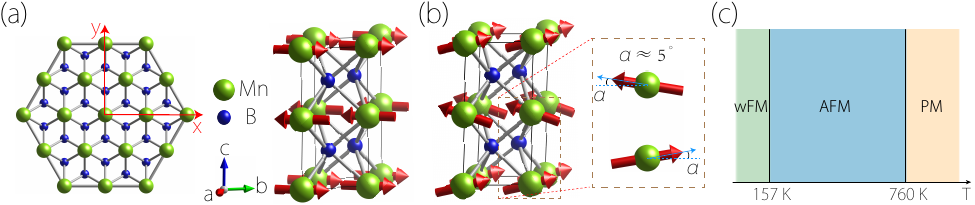}}
	\caption{\label{fig1} { (a) Top and side views of MnB$_2$ crystal. In the side view, we also illustrate the
AFM configuration. The red arrows show the direction of the local moment. (b) shows the wFM state of MnB$_2$.  The inset is an enlarged view indicating the tilting of magnetic moments towards $c$-axis (the angle is exaggerated here). (c) Magnetic phase diagram of MnB$_{2}$.} }
\end{figure*}

A recent focus in the field is to extend the study from nonmagnetic to magnetic systems. This is motivated by several points. First, due to magnetic ordering, symmetries of magnetic systems that underly topological semimetal states are different and are much richer compared to nonmagnetic systems. Second, magnetism offers a new possibility to control band topology and topological phase transitions, e.g., by tuning the magnetic moment orientation. Finally, magnetic materials are technologically important. Topological features and emergent fermions may endow them with new advantages for applications.
Currently, the realization of unconventional fermions in magnetic materials is still rather limited.  In Ref.~\cite{schoop2018tunable}, a ENP was reported in antiferromagnetic CeSbTe, but its energy is quite far away from Fermi level. Thus, it remains an important task to search for suitable magnetic materials that can host ENPs and other interesting topological states.

In this work, based on first-principles calculations and symmetry analysis, we reveal that the existing intermetallic compound MnB$_2$ is actually a magnetic topological semimetal with rich emergent fermion states. MnB$_2$ itself has two notable features~\cite{legrand1972neutron}. First, it has a high  N\'{e}el temperature of 760 K. Second,
it exhibits more than one magnetic phases. Between 157 K and 760 K, it adopts collinear antiferromagnetic (AFM) ordering.
Below 157 K, it transitions into a phase exhibiting weak ferromagnetism {(wFM)}, due to small tilting of the local moments.
We find that in the collinear AFM phase, MnB$_2$ possesses a pair of ENPs and an interesting nodal network formed by three sets of nodal loops in its low-energy band structure, protected by the spin group in the absence of spin-orbit coupling (SOC).
Depending on the N\'{e}el vector orientation, SOC will either open small gaps at these  degeneracies or transform them into
fourfold linear and quadratic Dirac points and separated nodal rings. Effective models are constructed for these nodal features. In addition, we find that due to symmetry constraint, there will a drastic change in the anomalous Hall response
accompanying the transition at 157 K: Anomalous Hall response is forbidden by symmetry in the collinear AFM phase, whereas it can become sizable in the wFM phase. We estimate that the intrinsic anomalous Hall conductivity can reach up to
$\sim 200$ $\Omega\cdot \text{cm}^{-1}$ below 157 K. Our work identifies a promising platform to study novel emergent fermions and their interplay with magnetic ordering.

\section{Computation Method}
Our first-principles calculations were based on the density functional theory (DFT), using a plane-wave basis set and projector augmented wave method~\cite{PhysRevB.50.17953}, as implemented in the Vienna \emph{ab} \emph{initio} simulation package~\cite{PhysRevB.54.11169,kresse1999ultrasoft}. The generalized gradient approximation (GGA) parameterized by Perdew, Burke, and Ernzerhof (PBE) was adopted for the exchange-correlation functional \cite{PhysRevLett.77.3865}. The energy cutoff was set to 390 eV. A $15\times15\times7$ Monkhorst-Pack $k$ mesh was used for the Brillouin zone (BZ) sampling. The atomic positions were fully optimized until the residual forces were less than $10^{-3}$ eV/{\AA}. The convergence criterion for the total energy was set to be $10^{-6}$ eV. To account for the correlation effects of the Mn-3\emph{d} electrons, we have adopted the GGA$+U$ method~\cite{PhysRevB.57.1505}. We have tested different $U$ values, and found that {$U=1$ eV gives a calculated magnetic moment $\sim 3  \mu_B$,  which best matches previous experimental observation ($\sim 2.6  \mu_B$)}~\cite{legrand1972neutron}, therefore, this $U$ value was taken for the results presented below. The $k\cdot p$ effective models were constructed with the help of the MagneticKP package~\cite{zhang2023magnetickp}. The intrinsic anomalous Hall conductivity was evaluated on a denser \textit{k} mesh of $200\times200\times200$ by using the WANNIER90 package~\cite{PhysRevB.74.195118,mostofi2008wannier90}.

\section{Crystal structure and magnetism}
The intermetallic compound MnB$_2$ has been synthesized in experiment more than sixty years ago~\cite{binder1960manganese}. The crystal has the hexagonal AlB$_2$ type structure with space group $P6/{mmm}$ (No.~191). As depicted in Fig.~\ref{fig1}, it consists of alternating atomic layers of Mn and B along the $c$ axis ($z$ direction in our setup).
Mn layers are of simple hexagonal type, whereas B layers are of honeycomb type.
A primitive unit cell contains one formula unit. The Mn atom occupies the 1\emph{a} Wyckoff position, whereas the B atoms are at the 2\emph{d} position. The experimental lattice constants are $a=3.01$ {\AA} and  $c=3.04$ {\AA}~\cite{cadeville1966proprietes}, which have been adopted in our first-principles calculations.

The magnetic structures of MnB$_2$ have been studied in previous experiments~\cite{legrand1972neutron}.
From magnetic measurements, such as nuclear magnetic resonance and neutron diffraction, it was found that MnB$_2$ has two magnetic phases. The magnetic moments are mainly on the Mn sites and have a value $\sim 2.6 \mu_B$.  At room temperature, MnB$_2$ has A-type collinear AFM, where each Mn atom layer is FM ordered and neighboring layers are coupled in AFM manner, as illustrated in  {{Fig.~\ref{fig1}(a)}.}
Measurement showed that the N\'{e}el vector perfers to be in the $ab$ plane. It was found that this collinear AFM phase occupies a wide temperature range from 157 K to 760 K. Below 157 K, MnB$_2$ enters the wFM phase, where the magnetic moments slightly tilt out of the plane, with a tilt angle $\sim 5^{\circ}$, as illustrated in {{Fig.~\ref{fig1}(b)}}, which leads to a network magnetization along $z$.  These experimental results will be used in our modeling below.


\begin{table*}[tp!]
	\centering
	\caption{{{Comparison of total energies for different magnetic configurations of MnB$_{2}$. Here the energies are with reference to the AFM-$y$ state and have unit of meV per formula unit.}}}
	\label{table1}
	\renewcommand\arraystretch{1.4}
	
	\begin{tabular}{p{0cm}<{\centering}p{2.5cm}<{\centering}p{2.5cm}<{\centering}p{2.5cm}<{\centering}p{2.5cm}<{\centering}p{2.5cm}<{\centering}p{2.5cm}<{\centering}p{2.5cm}<{\centering}}
		\hline\hline
		\rule{0pt}{13pt}
		&  Configuration &  AFM-$y$    & AFM-$x$   &AFM-$z$   &FM-$y$  \\
		\hline
		&  Total Energy &0.00  &0.27 & 0.40  &0.85 \\
		\hline\hline
	\end{tabular}
	\renewcommand\arraystretch{1.4}
\end{table*}

\section{Nonrelativistic band structure of AFM phase}
Let's investigate the electronic properties of MnB$_2$ in the collinear AFM phase [as in {{{Fig.~\ref{fig1}(a)}}].
We first consider the nonrelativistic band structure, i.e., in the absence of SOC. The result obtained from our DFT calculation is plotted in {{Fig.~\ref{fig2}(a)}.} One observes that
the system is metallic, which agrees with the transport measurement result reported in Ref.~\cite{kasaya1970magnetic} The density of states (DOS) shows a dip at the Fermi level, reflecting its semimetal character.
From projected DOS (PDOS), the low-energy bands are mainly contributed by Mn-3\emph{d} and B-2\emph{p} orbitals.

Interestingly, one observes several linear band crossings in the vicinity of Fermi level.
First of all, there is a crossing point $E$ on the $\Gamma$-$A$ path, located at $k_z=0.274\pi$ (in unit of $1/(2c)$, $2c$ is the lattice constant for the magnetic cell) and at energy of {{$-0.2575$ eV}} below Fermi level, as indicated by the red arrow in {{Fig.~\ref{fig2}(a)}} (There is also a partner point at $-0.274\pi$  due to inversion symmetry). Note that due to AFM ordering, spin degree of freedom has to be considered, and each band in the band structure is at least doubly degenerate with the spin degeneracy. Then, one can see from {Fig.~\ref{fig2}(a)} that the two bands which cross at $E$ are each fourfold degenerate, hence point $E$ is an ENP.


\begin{figure}[tb!]
	{\includegraphics[clip,width=8.2cm]{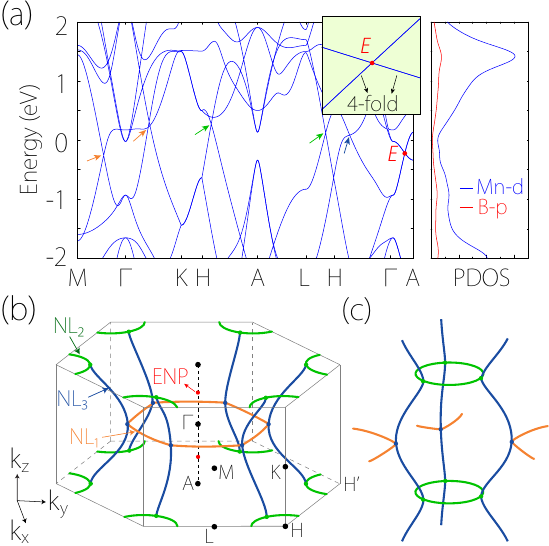}}
	\caption{\label{fig2}
		{ (a) Calculated band structure of MnB$_2$ along with PDOS, in the absence of SOC. The inset shows the enlarged view of the bands around ENP. (b) Distributions of the nodal lines, labeled as NL$_{1}$, NL$_2$, NL$_3$, in the BZ. They are connected and form a nodal network as shown in (c).
	}}
\end{figure}

Meanwhile, there are other crossing points marked by arrows with different colors in {{Fig.~\ref{fig2}(a)}}, a careful scan of BZ shows that these points actually belong to several nodal loops. As illustrated in {{Fig.~\ref{fig2}(b)}}, there is a nodal ring centered at $\Gamma$ point in the $k_z=0$ plane ({orange} colored in {{Fig.~\ref{fig2}(b)}}), marked as NL$_1$. Two (green colored) nodal rings are centered at $H$ and $H'$ points in the $k_z=\pi$ plane, marked as NL$_2$. In addition, there are six (blue colored) loops traversing BZ in the $z$ direction, marked as NL$_3$. These nodal loops are fourfold degenerate and they are connected to form an interesting network pattern as shown in {{Fig.~\ref{fig2}(c)}} in the extended BZ.

All these nodal features in the nonrelativistic band structure are protected by spin group symmetry~\cite{litvin1974spin}. Let's first consider the protection of the nodal loops. Without SOC, the spin-up and spin-down channels can be considered separately (and the two are connected, e.g., by translation of $c$ along $z$, which enforces the band spin degeneracy). If we focus on one spin channel, it can be regarded as a spinless system, which has an effective time reversal symmetry $T$ (represented by the complex conjugation). In addition, the inversion $P$ (with inversion center on Mn site), horizontal mirror $M_z$ (coincides with Mn plane), and three vertical mirrors are respected in each spin channel. These symmetries are sufficient to protect
the nodal loops in {{Fig.~\ref{fig2}(b)}}. For example, the loops in the $k_z=0$  and $k_z=\pi$ planes are protected by the $M_z$ symmetry, i.e., in each spin channel, the two crossing bands have opposite $M_z$ eigenvalues. Similarly, the six blue colored nodal loops are protected by the three vertical mirrors. Furthermore, one notes that each spin channel respects the spinless $PT$ symmetry, which enforces quantization of Berry phases in unit of $\pi$. This offers a second protection of the loops, as any small closed path encircling a loop here must carry a $\pi$ Berry phase. It follows that even if the mirrors are weakly broken, these loops will not disappear as along as the inversion symmetry is still preserved.


Now we turn to ENP (point $E$) and analyze its protection. In this analysis, we shall also construct an effective $k\cdot p$ model for its description. For spin groups, the actions on spatial and spin spaces are independent and can be separated.
We first consider the symmetries acting only on spatial (orbital) degrees of freedom. The ENP is formed by the crossing by two degenerate bands on $\Gamma$-$A$ path. For orbital degrees of freedom, there are two symmetry generators on this path, namely the sixfold rotation $C_{6z}$ and the vertical mirror $M_x$. The corresponding little co-group has two (single valued) two-dimensional (2d) irreducible representations (IRRs) $\Delta_5$ and $\Delta_6$ on this path.
The matrix representations for the generators in $\Delta_5$ and $\Delta_6$ may be taken as
\begin{equation}
	\begin{split}
	\Delta_5(C_{6z})=\begin{pmatrix}
		 -\frac{1}{2} & -\frac{\sqrt{3}}{2} \\
		\frac{\sqrt{3}}{2} & -\frac{1}{2} \\
	\end{pmatrix};\qquad
\Delta_5(M_x)= \begin{pmatrix}
1  & 0 \\
0 & -1 \\
\end{pmatrix},
\end{split}
\end{equation}
and
\begin{equation}
	\begin{split}
	\Delta_6(C_{6z})=\begin{pmatrix}
 \frac{1}{2} & -\frac{\sqrt{3}}{2} \\
\frac{\sqrt{3}}{2} & \frac{1}{2} \\
\end{pmatrix};\qquad
\Delta_6(M_x)=\begin{pmatrix}
	-1  & 0 \\
	0 & 1 \\
\end{pmatrix}.
	\end{split}
\end{equation}
Each 2D IRR gives a doubly degenerate band, and the accidental crossing between them will lead to a fourfold degenerate point on $\Gamma$-$A$.

{Next, we take into account the spin degree of freedom, which will double the degeneracy.
This introduces additional generators for the spin group. These include, e.g., all possible rotations of spin along its polarization direction. Assuming the moments are in the $\pm y$ direction (the specific direction does not matter here since there is no SOC), then such rotations, sometimes denoted as $S_y^\infty$~\cite{litvin1974spin}, can be represented as $\exp(-\frac{i\varepsilon \sigma_2}{2})$, where $\varepsilon$ is any real number and $\sigma_2$ is the Pauli matrix. In addition, reversing the spin direction
(e.g., by $\pi$ rotation normal to the spin polarization direction or by time reversal $\mathcal{T}$~\footnote{Here, we use script font $\mathcal{T}$ to distinguish it from the effective $T$ discussed above that does not acting on spin.}) and interchange the two AFM sublattices (e.g., by $P$ or by translation of $c$ along $z$) is also an allowed symmetry.
Including spin, $\Delta_n$ $(n=5,6)$ gives rise to spin group IRR $\Delta_n^s$ which is four dimensional, for which the generators can be taken as

\begin{equation}
	\begin{split}
		\Delta_n^s(C_{6z})=&\Delta_n(C_{6z}) \otimes \sigma_0,\qquad
		\Delta_n^s(M_x)=\Delta_n(M_x)\otimes \sigma_0,\\
		\Delta_n^s(S_{y}^{\infty})=&\sigma_0 \otimes e^{-\frac{i\varepsilon \sigma_2}{2}},\qquad
		\Delta_n^s(P\mathcal{T})=\sigma_0 \otimes i\sigma_2, \\
        \Delta_n^s(t_c||S_x^\pi)=&\sigma_0 \otimes e^{-\frac{i\pi\sigma_1}{2}},\qquad
        \Delta_n^s(t_c||S_z^\pi)=\sigma_0 \otimes e^{-\frac{i\pi\sigma_3}{2}},
	\end{split}
\end{equation}
where $\sigma_0$ is the $2\times 2$ identity matrix, and $t_c$ is the translation of $c$ along the $z$ direction (i.e., half of the magnetic cell length along $z$).

Using these matrix representations, we can construct the  $k\cdot p$ model for the ENP. Up to linear order in $q$ (i.e. the momentum deviation from point $E$), the model is given by
\begin{equation}
	\begin{aligned}
		\mathcal{H}_\text{ENP}(\bm q)=a_1 q_z +a_2 q_x \Gamma_{100} +{a_2} q_y\Gamma_{220}+a_3 q_z \Gamma_{300}.
	\end{aligned}
\end{equation}
Here, the model is an $8\times 8$ matrix with $\Gamma_{ijk}=\sigma_i \otimes \sigma_j \otimes \sigma_k$, $a$'s are real model parameters, and the energy is measured from the ENP.

Before proceeding, we comment that besides the nonmagnetic versus magnetic difference, all previously reported ENPs in fact require protection by nonsymmorphic space group symmetries and they all locate at high-symmetry points at the boundary of BZ~\cite{PhysRevLett.116.186402,doi:10.1126/science.aaf5037,rong2023realization}. In contrast, the ENP here is stabilized only by symmorphic symmetries and it occurs on a high-symmetry path in the interior of BZ. It also follows that the ENP here comes in a pair, whereas previous cases all have a single point.


\section{Effect of Spin-orbit coupling}
In this section, we examine the effect of SOC in the collinear AFM phase. With SOC, the orientation of N\'{e}el vector will affect band structure.  Note that the previous experiments only showed the N\'{e}el vector prefers to be in-plane, but they did not determine the specific in-plane direction~\cite{legrand1972neutron}. Our DFT calculation suggests that the energy of $y$ direction is slightly lower than the $x$ direction (see {{Table~I}}). (The qualitative features of results for the two in-plane directions are essentially the same.)

Let's first consider the case with N\'{e}el vector along the $y$ direction, denoted as AFM-$y$ configuration. With SOC, symmetry operations will act on orbital and spin spaces simultaneously. Note that the system still respects $P\mathcal{T}$ symmetry, where the inversion center is in the B layer at the center of a B hexagon. As a result, each band still has a twofold spin degeneracy due to $(P\mathcal{T})^2=-1$. Furthermore, we find that SOC generally opens a small gap at all the nodal features near Fermi level. Figure{{~\ref{fig3}(a) and \ref{fig3}(b)}}  show the enlarged band structures around two original band crossings on high-symmetry paths. The opened SOC gap ranges from a few meV to {$\sim 8$} meV.


\begin{figure}[tb!]
	{\includegraphics[clip,width=8.2cm]{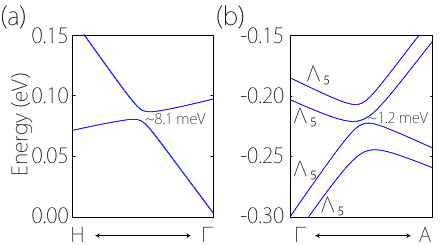}}
	\caption{\label{fig3} {Band structure of MnB$_2$ with SOC included along (a) $H$-$\Gamma$ and (b) $\Gamma$-$A$ paths for the AFM-$y$ state. Small gaps are opened at the original band crossings by SOC. }}
\end{figure}

Next, we consider the configuration with N\'{e}el vector along the $z$ direction, denoted as AFM-$z$. In this case, we find that there are still preserved nodal features near Fermi level. First of all, regarding the nodal network in {{Fig.~\ref{fig2}(b)}}, we find that the two nodal rings in the $k_z=\pi$ plane (the NL$_2$ rings) are preserved, while other nodal lines open small gaps, as illustrated in {{Fig.~\ref{fig4}(a)}}. The symmetry protection of these fourfold nodal rings come from the combined actions of the horizontal mirror $M_z$ and $P\mathcal{T}$ symmetries. Here, each nodal ring is formed by the crossing of two doubly degenerate bands, where the double degeneracy comes from $P\mathcal{T}$. To have a protecting crossing, it is essential that the $P\mathcal{T}$ related partners must have the same $M_z$ eigenvalues. To demonstrate this point, consider a band eigenstate $|\psi\rangle$ with momentum in the $k_z=\pi$ plane. It can always be chosen as an $M_z$ eigenstate, i.e.,
\begin{equation}
  M_z|\psi\rangle=m_z|\psi\rangle,
\end{equation}
where $m_z=i$ or $-i$ in the presence of SOC. To find the $M_z$ eigenvalue of its partner $P\mathcal{T}|\psi\rangle$, we need to know the relationship between operations $M_z$ and $P\mathcal{T}$. Considering their consecutive actions on a spatial point $(x, y, z)$, we have

{
\begin{equation}
	(x,y,z)\xrightarrow[]{\mathcal{PT},M_z}\left(-x,-y,z+\frac{1}{2}\right),
\end{equation}
and
\begin{equation}
	(x,y,z)\xrightarrow[]{M_{z}, \mathcal{PT}}\left(-x,-y,z-\frac{1}{2}\right).
\end{equation}
}

Therefore, we have
\begin{equation}\label{mz}
  M_z (P\mathcal{T}|\psi\rangle)=e^{-ik_z}P\mathcal{T}M_z|\psi\rangle=m_z(P\mathcal{T}|\psi\rangle),
\end{equation}
where in the second step, we have used $k_z=\pi$ and $m_z=\pm i$ is purely imaginary. This demonstrates that the $P\mathcal{T}$ partners $|\psi\rangle$ and $P\mathcal{T}|\psi\rangle$ must have the same eigenvalue $m_z$. When bands with opposite $m_z$ values cross, they will form the fourfold band crossing (see {{Fig.~\ref{fig4}(a)}}), i.e., the NL$_2$ rings. One also notes from Eq.~(\ref{mz}) that
the condition $k_z=\pi$ is important; the other mirror plane $k_z=0$ does not have this property. This explains why the original nodal ring NL$_1$ in {{Fig.~\ref{fig2}(a)}} is not preserved.

Next, we consider the SOC effect on the ENP in the AFM-$z$ state. We find that each ENP will be transformed into three fourfold Dirac points on $\Gamma$-$A$ path. These points are marked in the calculated band structure in {{Fig.~\ref{fig4}(c)}}. Two points are conventional linear Dirac points (marked as $D_1$ and $D_2$ in the figure). Interesting, the third one is a quadratic Dirac point~\cite{PhysRevB.98.125104,PhysRevB.101.205134} (marked as $D_3$), which has linear band splitting along $k_z$ but quadratic band splitting in the 2D plane normal to $k_z$.

\begin{figure}[tb!]
	{\includegraphics[clip,width=8.2cm]{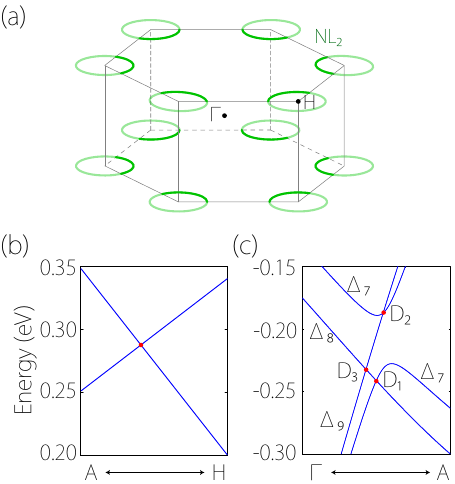}}
	\caption{\label{fig4} {Band structure results for AFM-$z$ state under SOC. (a) Two nodal rings (the NL$_2$ rings) in $k_z=\pi$ plane are still preserved under SOC. (b) Band structure along $A$-$H$ path, showing the fourfold degenerate point on the nodal ring.  (c) Band structure around the original ENP on $\Gamma$-$A$. Here, $D_1$ and $D_2$ are linear Dirac point, whereas $D_3$ is a quadratic Dirac point.}}
\end{figure}

To characterize these Dirac points, we construct their $k\cdot p$ effective models.
The magnetic space group for AFM-\emph{z} is $P_c6/mmc$ (BNS No.~192.252).
The low-energy bands that form the Dirac points belong to $\Delta_7$, $\Delta_8$, and  $\Delta_9$ irreducible representations on the $\Gamma$-$A$ path, as shown in {{Fig.~\ref{fig4}(c)}}. The symmetry generators on this path can be selected as
$C_{6z}$, $\tilde{M}_x$, and $P\mathcal{T}$, where $\tilde{M}_x$ is a glide mirror involving a half magnetic cell translation along $z$. Their matrix representations for $\Delta_7$, $\Delta_8$, and  $\Delta_9$ can be taken as
%
%
%
%
%
\begin{equation}
	\begin{split}
		\Delta_7(C_{6z})=&\begin{pmatrix}
			i  & 0 \\
			0 & -i \\
		\end{pmatrix},\qquad
		\Delta_8(C_{6z})=\begin{pmatrix}
			e^{\frac{i\pi }{6}} & 0 \\
			0 & e^{-\frac{i\pi }{6}}\\
		\end{pmatrix},\\
		\Delta_9(C_{6z})=&\begin{pmatrix}
	e^{\frac{i 5\pi }{6}} & 0 \\
	0 & e^{-\frac{i 5\pi }{6}}\\
\end{pmatrix},
\end{split}
\end{equation}
and
\begin{equation}
		\Delta_n(\tilde{M}_x)=e^{-ik_z}\begin{pmatrix}
	0  & i \\
	i & 0 \\
\end{pmatrix};
\end{equation}
\begin{equation}
		\Delta_n(P\mathcal{T})=\begin{pmatrix}
		0  & 1 \\
		-1 & 0 \\
	\end{pmatrix},
\end{equation}
for $n=7, 8, 9$. For the linear Dirac point $D_1$ formed by the crossing of $\Delta_7$ and $\Delta_8$ bands, the obtained effective model takes the form of
\begin{equation}
	\begin{aligned}
		\mathcal{H}_{D_1}(\bm q)&=b_1 q_z +b_2 q_x \Gamma_{23}+ b_2 q_y \Gamma _{10}+ b_3 q_z \Gamma _{30},
	\end{aligned}
\end{equation}
where $\Gamma_{ij}=\sigma_i\otimes \sigma_j$. Similar, point $D_2$ formed by the  crossing of $\Delta_7$ and $\Delta_9$
is described by the model of
\begin{equation}
	\begin{aligned}
		\mathcal{H}_{D_2}(\bm q)&=c_1 q_z +c_2 q_x \Gamma_{23}-c_2 q_y \Gamma _{10}+ c_3 q_z \Gamma _{30}.
	\end{aligned}
\end{equation}
Lastly, the quadratic Dirac point $D_3$ is formed by the crossing of $\Delta_8$ and $\Delta_9$. Its model is given by
\begin{equation}
	\begin{aligned}
		\mathcal{H}_\text{QDP}(\bm q)=& d_1 q_z+d_2 (q_x^2+q_y^2)+d_3 q_z\Gamma_{30}+ d_4(q_x^2-q_y^2)\Gamma_{10}\\
&+2d_4 q_x q_y\Gamma_{23}
+d_5(q_x^2+q_y^2)\Gamma_{30}.
	\end{aligned}
\end{equation}
In the above models, the energy and the momentum are measured from each nodal point, and $b_i$'s, $c_i$'s, and $d_i$'s are real model parameters. These models confirm that $D_1$ and $D_2$ are linear Dirac points, and $D_3$ is a quadratic Dirac point, all realized in an AFM state. It should be noted that although quadratic Dirac point was proposed before based on theoretical analysis in nonmagnetic and magnetic systems~\cite{PhysRevB.98.125104,PhysRevB.101.205134}. Its material realization is rare. Particularly, to our knowledge, it has not been reported in magnetic materials before.


%
%
%



\section{Anomalous Hall response in FM phase}

Experiment shows that when the temperature is below 157 K, MnB$_2$ will transition  into the wFM phase, with magnetic moments  tilted toward the \emph{c} axis by a small angle ($\sim 5^{\circ}$)~\cite{legrand1972neutron}, as shown in {{Fig.~\ref{fig1}(b)}}. The magnetic space group changes into {{$Cm'c'm$ (BNS No.~63.462)}}, in which the symmetries $P$ and $M_z$ are respected. The calculated band structure of the wFM phase
in shown in {{Fig.~\ref{fig5}(a)}}. One observes that, first, the twofold band degeneracy in the AFM phase is lifted in the wFM phase.
Second, the system is still metallic, and low-energy bands show complicated crossing patterns.

Here, we are interested in the anomalous Hall response. The reason is that in the AFM phase, regardless of the N\'{e}el vector orientation, this response is forbidden in bulk MnB$_2$, by the $\mathcal{T}t_{00\frac{1}{2}}$ symmetry, where $t_{00\frac{1}{2}}$ is the translation along $z$ by half of the AFM cell. Now, in the wFM phase, this $\mathcal{T}t_{00\frac{1}{2}}$ symmetry is broken by the moment tilting, so the system can exhibit a finite anomalous Hall response. This can serve as a contrasting property between the two phases.




\begin{figure}[tb!]
	{\includegraphics[clip,width=8.2cm]{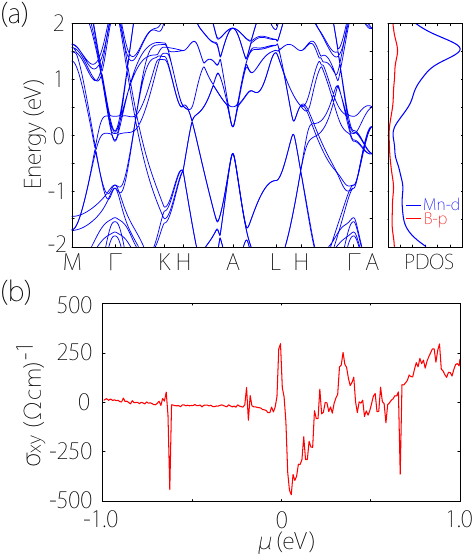}}
	\caption{\label{fig5}{(a) Band structure for wFM state along with PDOS. (b) Calculated intrinsic anomalous Hall conductivity as a function of chemical potential for the wFM state.}}
\end{figure}
We evaluate the intrinsic part of anomalous Hall conductivity in the $ab$ plane, which is related to the
Berry curvature of the material's band structure~\cite{PhysRevLett.88.207208,RevModPhys.82.1539}:
\begin{equation}
\sigma_{x y}=-\frac{e^{2}}{\hbar} \int_{\mathrm{BZ}} \frac{d^{3} k}{(2 \pi)^{3}} \Omega_{z}(\boldsymbol{k}).
\end{equation}
Here, $\Omega_{z}(\bm k)$ is the total Berry curvature of all occupied states at $\bm k$
\begin{equation}
\Omega_{z}(\boldsymbol{k})=-2 \operatorname{Im} \sum_{n \neq n^{\prime}} f_{n \boldsymbol{k}} \frac{\left\langle n \boldsymbol{k}\left|v_{x}\right| n^{\prime} \boldsymbol{k}\right\rangle\left\langle n^{\prime} \boldsymbol{k}\left|v_{y}\right| n \boldsymbol{k}\right\rangle}{\left(\omega_{n^{\prime}}-\omega_{n}\right)^{2}}
\end{equation}
where $n$ and $n'$ are band indices, $\varepsilon_{n}=\hbar \omega_{n}$ is the band energy, $v$'s are the velocity operators, and $f_{n \boldsymbol{k}}$ is the Fermi distribution function for state $|n\bm k\rangle$.

Figure~{{\ref{fig5}(b)}} shows the $\sigma_{xy}$ obtained from our DFT calculations as a function of the chemical potential $\mu$.
We find that without doping ($\mu=0$), $\sigma_{xy}$ has a value of $\sim200$ $(\Omega \cdot \text{cm})^{-1}$, which is actually sizable, considering that good FM materials such as Fe has a value around $\sim 1000$ $(\Omega \cdot \text{cm})^{-1}$~\cite{PhysRevLett.92.037204}. Moreover, with small electron doping, $\sigma_{xy}$ can have a large change and even flip its sign, reaching $\sim-450$ $(\Omega \cdot \text{cm})^{-1}$ at $\mu\sim 60$ meV (i.e., doping of $\sim 0.018$ e/f.u.).
These results show that the anomalous Hall signal will exhibit a large jump during the magnetic phase transition, from nearly zero in the AFM phase to a sizable value in the wFM phase.



\section{Discussion and Conclusion}
We have revealed interesting band nodal features in MnB$_2$. The ENPs found here is distinct from previously reported cases, which are either in nonmagnetic systems or require nonsymmorphic symmetries~\cite{PhysRevLett.116.186402,doi:10.1126/science.aaf5037,rong2023realization}. As we discussed, the ENPs here are realized in AFM state and requires only symmorphic spin group symmetries. It follows that they appear in a pair, rather than a single point in previous examples. The AFM quadratic Dirac point is another interesting discovery. It should be noted that all Dirac points here have zero Chern numbers, because the system has both $\mathcal{T}t_{00\frac{1}{2}}$ and $P$ symmetries.

Experimentally, single crystal MnB$_2$ has already been synthesized~\cite{binder1960manganese}. The nodal features predicted here can be imaged by the angle-resolved photoemission spectroscopy (ARPES) technique. Note that the AFM phase exists between 157 K and 760 K. In this range, the SOC induced splitting in low-energy bands are comparable or even less than the temperature broadening. Hence, the small SOC gap may not be resolved in ARPES images, so nodal features like ENP and nodal network can be well perceived. Finally, the anomalous Hall response can be studied with the standard Hall bar configuration.

In conclusion, we discover topological semimetal states in an existing magnetic material MnB$_2$. In the AFM phase above 157 K, we reveal a pair of ENPs coexisting with a nodal network. The ENPs are protected by symmorphic spin group symmetries and located on a high-symmetry path, distinct from previous cases. The nodal network is composed of three types of nodal loops, and each loop enjoys a double symmetry protection. SOC may open a small gap at these degeneracies. In the  AFM-\emph{z} state,
a pair of nodal rings are robust against SOC, and the ENP splits into three Dirac points, one of which is a magnetic quadratic Dirac point. For the wFM phase below 157 K, we predict a sizable anomalous Hall response $\sim200$ $(\Omega \cdot \text{cm})^{-1}$, which contrasts to the AFM phase with the response forbidden by symmetry. Our work reveals a concrete material platform for studying magnetic topological semimetal states with unconventional emergent fermions. The jump in anomalous Hall signal could be used to probe the magnetic transition and may be useful for designing new spintronic devices.


\begin{acknowledgements}
	The authors thank D. L. Deng for valuable discussions. This work is supported by the Project of Educational Commission of Hunan Province of China (Grant No. 21A0066), the Hunan Provincial Natural Science Foundation of China (Grant No. 2022JJ30370), the National Natural Science Foundation of China (Grants No. 11704117), the UGC/RGC of Hong Kong SAR (AoE/P-701/20) and Singapore NRF CRP (CRP22-2019-0061). We acknowledge computational support from H2 clusters in Xi'an Jiaotong University and National Supercomputing Centre Singapore.
\end{acknowledgements}

\bibliography{afm_refs}

\bibliographystyle{apsrev4-1}

\end{document}